\documentclass[conference]{IEEEtran}
\IEEEoverridecommandlockouts
\usepackage{cite}
\usepackage{pifont}
\usepackage{amsmath,amssymb,amsfonts}
\usepackage{algorithmic}
\usepackage{graphicx}
\usepackage{textcomp}
\usepackage[table,xcdraw]{xcolor}
\usepackage{booktabs}
\usepackage{tablefootnote}
\usepackage{xcolor}
\usepackage[T1]{fontenc}    % use 8-bit T1 fonts
\def\BibTeX{{\rm B\kern-.05em{\sc i\kern-.025em b}\kern-.08em
    T\kern-.1667em\lower.7ex\hbox{E}\kern-.125emX}}
\usepackage{multirow}

\usepackage{hyperref}       % hyperlinks
\hypersetup{
    colorlinks=true,
    linkcolor=blue,
    filecolor=magenta,      
    urlcolor=black,
    citecolor=magenta,
    pdftitle={Overleaf Example},
    pdfpagemode=FullScreen,
}

\newcommand{\cmark}{\ding{51}}%
\newcommand{\xmark}{\ding{55}}%

\begin{document}

% \title{Improving Speech Enhancement with A Holistic Speech Quality Predictor\\
% \title{Exploring Holistic Quality Supervision for Speech Enhancement via Learned Quality Assessment
\title{Improving Speech Enhancement with Multi-Metric Supervision from Learned Quality Assessment
}

% \author{\IEEEauthorblockN{Anonymous submission to ASRU2025}}
\author{\IEEEauthorblockN{Wei Wang\textsuperscript{1}, Wangyou Zhang\textsuperscript{1}, Chenda Li\textsuperscript{1}, Jaitong Shi\textsuperscript{2}, Shinji Watanabe\textsuperscript{2}, Yanmin Qian\textsuperscript{1}}
\IEEEauthorblockA{
\textit{
$^1$Shanghai Jiao Tong University, China
$^2$Carnegie Mellon University, USA}
}
}
% \IEEEauthorblockA{
% \textsuperscript{1}Auditory Cognition and Computational Acoustics Lab \\
% MoE Key Lab of Artificial Intelligence, AI Institute \\
% School of Computer Science, Shanghai Jiao Tong University, Shanghai, China\\
% \textsuperscript{2}School of Artificial Intelligence, Shanghai Jiao Tong University, Shanghai, China\\
% \textsuperscript{3}Language Technologies Institute, Carnegie Mellon University, PA, USA}
% }

\maketitle

\begin{abstract}
Speech quality assessment (SQA) aims to predict the perceived quality of speech signals under a wide range of distortions.  It is inherently connected to speech enhancement (SE), which seeks to improve speech quality by removing unwanted signal components. 
While SQA models are widely used to evaluate SE performance, their potential to guide SE training remains underexplored. In this work, we investigate a training framework that leverages a SQA model, trained to predict multiple evaluation metrics from a public SE leaderboard, as a supervisory signal for SE. This approach addresses a key limitation of conventional SE objectives, such as SI-SNR, which often fail to align with perceptual quality and generalize poorly across evaluation metrics. Moreover, it enables training on real-world data where clean references are unavailable. Experiments on both simulated and real-world test sets show that SQA-guided training consistently improves performance across a range of quality metrics. Code and checkpoints are available~\footnote{\url{https://github.com/urgent-challenge/urgent2026_challenge_track2}}.

\end{abstract}


\section{Introduction}
\label{sec:intro}

Speech enhancement (SE) aims to improve the quality and intelligibility of speech signals by removing background noise, reverberation, and other distortions~\cite{Speech-Loizou2013}. It plays a vital role in many real-world applications, including teleconferencing, hearing aid, and robust automatic speech recognition. Despite progress in deep learning-based SE methods, two major challenges persist. First, most SE models are trained using supervised objectives that require clean reference signals, such as scale-invariant signal-to-noise ratio (SI-SNR)~\cite{SDR__Half_baked-LeRoux2019} or spectral distance. This reliance on simulated paired data limits the ability to utilize real-world recordings and leads to a mismatch between training and inference~\cite{OA-Iwamato2022,BridgeTheGap-Wang2020,OA-Wang2025,FatHuBERT-Yang2023,ClosingTheGap-Zhang2021, LessIsMore-Li2025, zhao25h_interspeech}. Second, these low-level signal-based objectives do not always align with human perceptual preferences, often resulting in suboptimal mean opinion score~(MOS) when evaluated with subjective human listening tests~\cite{MOS-Cooper2024,MOS-Torcoli2021,VERSA-Shi2024,LessionsUrgent2024-Zhang2025}.

% Speech quality assessment (SQA) models, designed to predict the perceived quality of speech signals, are inherently connected to SE. Most existing SQA models are trained to estimate mean opinion scores (MOS), either from human-labeled data or by distilling judgments from more complex evaluation systems. For instance, DNSMOS~\cite{DNSMOS-Reddy2022} focuses on denoised speech using a lightweight model, UTMOS~\cite{UTMOS-Saeki2022,UTMOSv2-Baba2024} is trained for broader generalization across conditions, and SIGMOS~\cite{SIGMOS-Ristea2025} decomposes quality into interpretable subcategories such as noise, reverberation, and signal integrity. While these models effectively capture specific perceptual dimensions, they generally focus on a limited output metrics. To enable more comprehensive quality modeling, Uni-VERSA~\cite{Uni-VERSA-Shi2025} was recently proposed as a multi-metric SQA model that jointly predicts a diverse set of metrics—including intelligibility, naturalness, fidelity, and speaker consistency—from a single input. Initially developed for post-hoc evaluation of SE systems, Uni-VERSA provides a unified interface for assessing multiple quality dimensions simultaneously.

Speech quality assessment (SQA) has emerged as a vibrant area of research, driven by the growing demand for automated, perceptually aligned evaluation tools. The recent introduction of the Audiobox Aesthetics~\cite{Audiobox-Tjandra2025} underscores this trend by providing a unified framework for assessing the aesthetic quality of diverse audio modalities, including speech, music, and environmental sounds. In the field of SE, SQA models~\cite{DNSMOS-Reddy2022,UTMOS-Saeki2022, UTMOSv2-Baba2024,VoiceMos2022-Huang2022,VoiceMos2023-Cooper2023,VoiceMos2024-Huang2024, UrgentPK-Wang2025} are typically designed to estimate MOS either from human-labeled datasets or by distilling knowledge from complex evaluation systems. Notable examples include DNSMOS~\cite{DNSMOS-Reddy2022}, a lightweight model targeting denoised speech, and UTMOS~\cite{UTMOS-Saeki2022,UTMOSv2-Baba2024}, which aims to generalize across broader acoustic conditions, and SIGMOS~\cite{SIGMOS-Ristea2025} decomposes quality into interpretable subcategories such as noise, reverberation, and signal integrity. While effective, these models often focus on a single or limited set of perceptual metrics. To support richer, multi-dimensional supervision in SE, Uni-VERSA~\cite{Uni-VERSA-Shi2025} was recently proposed as a multi-metric SQA model. It jointly predicts a diverse set of perceptual metrics—including intelligibility, naturalness, fidelity, and speaker consistency—from a single speech input. Initially developed for post-hoc evaluation of SE systems, Uni-VERSA provides a unified interface for assessing multiple quality dimensions simultaneously.

In this work, we explore the use of a multi-metric SQA model to supervise SE training. We first address the limitations of the original Uni-VERSA model by expanding its training set to a larger and more diverse collection of enhanced utterances and increasing the number of predicted metrics from 11 to 22. The resulting model, referred to as Uni-VERSA-Ext, provides broader quality coverage and improved generalization across evaluation dimensions. Guiding SE training with such a learned multi-metric SQA model offers several key advantages. (1) Many widely used evaluation metrics (e.g. STOI, PESQ~\cite{PESQ-Rix2001} and POLQA~\cite{POLQA-Beerends2013}) are non-differentiable and therefore unsuitable as direct optimization targets. The SQA model serves as a differentiable proxy, allowing SE models to be optimized toward these otherwise inaccessible criteria. (2) Metrics such as speaker similarity and character error rate~(CER) require large and computationally expensive back-end models to compute, which significantly increases training cost and memory usage. The SQA model enables efficient supervision without incurring such overhead. (3) For real-world data without clean references, the SQA model makes it possible to train SE models using evaluation-aligned objectives, effectively mitigating the train–inference mismatch. 

We further explore score-based and feature-based objectives for integrating Uni-VERSA-Ext into SE training. When training on real-world audio, we observe that SE models can exploit weaknesses in the SQA model, producing adversarial outputs that receive high quality scores without corresponding perceptual improvements. To mitigate this, we introduce a self-supervised regularization term that constrains the enhanced outputs, stabilizes training, and prevents the SQA model from being applied to out-of-distribution or adversarial inputs. All training data used in our experiments are reproducible and will be made publicly available.

Our contributions are summarized as follows:
\begin{enumerate}
\item We introduce \textbf{Uni-VERSA-Ext}, an enhanced version of Uni-VERSA with improved generalization and an expanded output space covering 22 speech quality metrics.
\item We investigate several strategies for leveraging Uni-VERSA-Ext in SE training on both simulated and real-world data.
\item We analyze failure cases in SQA-guided fine-tuning on real-world data and propose a self-supervised regularization term to improve robustness on real-world audio.
\end{enumerate}

\section{Proposed Method}
\label{sec:proposed}

\subsection{Extension of the Uni-Versa SQA Model}

To enable perceptually aligned supervision for speech enhancement (SE) training, we build upon Uni-VERSA~\cite{Uni-VERSA-Shi2025}, a recently proposed multi-metric speech quality assessment (SQA) model. While Uni-VERSA provides a unified framework for evaluating various quality metrics, we find it necessary to enhance its generalization and robustness before using it for gradient-based supervision. We thus introduce \textbf{Uni-VERSA-Ext}, an extended version that shares the same model architecture as Uni-VERSA but incorporates three key modifications to address major limitations.

\paragraph{Expanded Training Data} The original Uni-VERSA was trained on approximately 100k enhanced utterances, all derived from just 850 source noisy recordings\footnote{The rest 150 source noisy recordings were held out for development and evaluation in our experiments.} in the blind test set of the URGENT 2024 challenge. Although the dataset includes both simulated and real recordings, its limited diversity in linguistic and acoustic content restricts the model’s generalization capability.

To overcome this, we expand the training set to over 500k utterances from 5,750 source recordings, covering submissions from both the URGENT 2024~\cite{URGENT-Zhang2024} and URGENT 2025~\cite{URGENT2025} challenges. This expansion introduces significantly more content and distortion variability, along with a greater amount of multilingual and MOS-labeled data.

% \paragraph{Extended Metric Space} The original Uni-VERSA predicts 11 quality metrics. To enable broader and more informative supervision, we expand the prediction space to 22 metrics. Table~\ref{tab:universa-extended} summarizes the set of evaluation metrics supported by Uni-VERSA and Uni-VERSA-Ext. Compared to the original Uni-VERSA, the extended version incorporates several important additions as discussed in Section~\ref{subsec:metrics_coverage}. 

\paragraph{Extended Metric Space} The original Uni-VERSA predicts 11 quality metrics. To enable broader and more informative supervision, we expand the prediction space to 22 metrics. Table~\ref{tab:universa-extended} summarizes the set of evaluation metrics supported by Uni-VERSA and Uni-VERSA-Ext. Compared to the original Uni-VERSA, the extended version incorporates several important additions as discussed in Section~\ref{subsec:metrics_coverage}. Additionally, we replace several metrics to improve robustness and reliability: SI-SNR is replaced by SDR due to its sensitivity to phase mismatch; WER is replaced by CER to better support multilingual speech; and STOI is replaced by ESTOI, which has been shown to correlate better with human perception, especially under high-SNR conditions.

% SIGMOS-based metrics are included to enable finer-grained assessment of noise and distortion levels. The evaluation of naturalness is improved through the inclusion of additional neural MOS predictors. Furthermore, a rank-based score from the URGENT benchmarks~\cite{URGENT-Zhang2024} is added to provide a relative perspective on overall quality. 

\paragraph{Output Range Stabilization} We observed that the original Uni-VERSA can produce invalid outputs, such as negative WER scores, which are not semantically meaningful. To ensure the validity and stability of predictions, we introduce metric-specific activation functions to the output layer. For example, we apply ReLU for metrics with non-negative ranges (e.g., CER), and scaled sigmoid functions for metrics bounded within intervals like [0, 1] or [1, 5]. These activations improve output interpretability and ensure compatibility with loss computation during SE training.

\vspace{-.5em}

\subsection{Metrics Coverage in Uni-VERSA-Ext}
\label{subsec:metrics_coverage}

Uni-VERSA-Ext spans multiple domains, as indicated in the \textit{Domain} column of Table~\ref{tab:universa-extended}, providing broader coverage than Uni-VERSA and facilitating improved generalization.

\paragraph{Noise \& Distortion Level}
This domain assesses perceived speech quality under various types of degradation. PESQ~\cite{PESQ-Rix2001} and SDR~\cite{SDR-Vincent2006} are included to measure perceptual quality and signal preservation. To enable fine-grained evaluation, we further incorporate \textit{DNSMOS}~\cite{DNSMOS-Reddy2022} (DNS. for short), which predicts overall quality scores, along with the SIGMOS~\cite{SIGMOS-Ristea2025} metric suite. \textit{SIGMOS\_OVRL} (SIG.OVRL for short) estimates overall quality, while \textit{SIGMOS\_NOISE}, \textit{REVERB}, \textit{COL}, \textit{LOUD}, and \textit{SIG} capture specific degradations such as background noise, reverberation, spectral coloration, loudness inconsistency, and signal artifacts. Together, these metrics provide a broad view of distortion effects in real-world and synthetic conditions.

\paragraph{Naturalness}
This domain captures how natural and human-like the speech sounds. Uni-VERSA-Ext employs multiple predictive models to approximate subjective listening scores. \textit{UTMOS}~\cite{UTMOS-Saeki2022} (UT. for short) serves as a general-purpose quality predictor trained on diverse data. \textit{Distill\_MOS}~\cite{Distill_MOS-Stahl2025}  (Distill. for short) is a compact variant distilled from human-annotated scores. \textit{NISQA\_MOS}~\cite{NISQA-Mittag2021}  (NISQA for short) focuses on speech processed by telecommunication and enhancement systems, while \textit{SCOREQ}~\cite{SCOREQ-Ragano2024} provides a unified quality estimate across domains. 

\paragraph{Intelligibility}
This domain reflects how clearly the linguistic content of speech is conveyed. We include \textit{CER} as a proxy for recognition performance, evaluated using the OWSM~\cite{OWSM-Peng2024} model\footnote{\url{https://huggingface.co/espnet/owsm_v3.1_ebf}}. 
\textit{ESTOI}~\cite{ESTOI-Jensen2016} captures intelligibility under adverse acoustic conditions by emphasizing short-time temporal dynamics. \textit{SpeechBERTScore}~\cite{SpeechBERTScore-Saeki2024} measures semantic alignment between utterances using embeddings from the multilingual HuBERT model\footnote{\url{https://huggingface.co/utter-project/mHuBERT-147}}~\cite{HuBERT-Hsu2021}. \textit{PhonemeSimilarity}~\cite{LPS-Pirklbauer2023} quantifies phonetic consistency based on the Levenshtein distance between phoneme sequences predicted by a phoneme recognition model\footnote{\url{https://huggingface.co/facebook/wav2vec2-lv-60-espeak-cv-ft}}.

\paragraph{Speaker Characteristics}
We evaluate speaker similarity by computing the cosine similarity between speaker embeddings extracted using a pretrained speaker verification model\footnote{\url{https://huggingface.co/espnet/voxcelebs12_rawnet3}}~\cite{ESPnet-SPK-Jung2024}. This metric assesses the consistency of speaker identity between reference and enhanced speech.

\paragraph{Spectral Accuracy}
We include \textit{mel-cepstral distortion (MCD)}~\cite{MCD-Kubichek1993}, a widely used metric in text-to-speech and voice conversion tasks, to quantify frame-level spectral deviations. MCD provides a fine-grained assessment of low-level acoustic fidelity by measuring how closely the generated speech matches the expected spectral structure.

\paragraph{Overall Quality}
We incorporate a ranking-based metric inspired by the evaluation protocol used in the URGENT challenges~\cite{URGENT-Zhang2024}. The ranking score is computed by first ranking enhanced samples for each metric and then aggregating the ranks using a weighted sum to produce an overall quality score. Unlike the original URGENT setup, which evaluates ranking at the system level, we adopt an utterance-level approach. Specifically, for each source recording, we collect all corresponding enhanced utterances, compute individual metric scores, and derive a rank-based score for each utterance. To ensure comparability across different source recordings, the final rank score is normalized by the number of enhanced versions per source. This metric has been shown in the URGENT challenge to correlate well with human MOS ratings~\cite{Lessons-Zhang2025}, making it a practical proxy for overall perceptual quality.

\begin{figure}[htbp]
\centerline{\includegraphics[width=\linewidth]{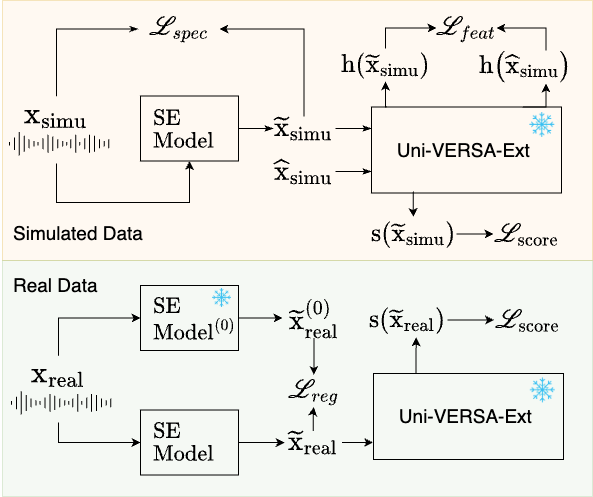}}
\caption{Overview of the training pipeline for SQA-guided SE using either real or simulated data.}
\label{fig:overview}
\vspace{-2em}
\end{figure}

\subsection{SQA Supervision with Predicted Score}

% We begin by using Uni-VERSA-Ext to supervise speech enhancement based on predicted quality scores.
We use Uni-VERSA-Ext to supervise SE via its predicted quality scores.
Let \( \mathbf{x} \) denote the input noisy speech signal, and \( \tilde{\mathbf{x}} \) denote the corresponding enhanced signal produced by the model under training. The enhanced signal is passed into the frozen Uni-VERSA-Ext model, which outputs both a set of quality scores and an internal hidden representation:
\begin{equation}
\left[\mathbf{s}, \mathbf{h} \right] = \text{Uni-VERSA-Ext}(\tilde{\mathbf{x}}),
\end{equation}
where \( \mathbf{s} = [s_1, s_2, \ldots, s_K] \) represents the predicted values of evaluation metrics, and \( \mathbf{h} \) denotes a latent feature vector from the penultimate layer of the SQA model.

Among the predicted scores, some metrics are quality-oriented (higher is better), while others are error-based (lower is better). We assign a coefficient \( \alpha_k \in \{-1, +1\} \) to each metric: \( \alpha_k = -1 \) for positively oriented metrics and \( \alpha_k = +1 \) for negatively oriented ones.

The score-based supervision loss is then computed as:
\begin{equation}
\mathcal{L}_{\text{score}} = \sum_{k=1}^{K} w_k \cdot \alpha_k \cdot s_k,
\end{equation}
where \( w_k \) is a metric-specific weight (defaulting to 1 unless otherwise specified). It is used for training on both real and simulated data as shown in Figure~\ref{fig:overview}.

\subsection{SQA Supervision in Feature Space}

While score-based supervision encourages improved quality scores, it may overlook the rich intermediate representations learned by the SQA model. To address this, we introduce a feature-space supervision strategy that leverages the hidden features extracted by the SQA model as training targets.

This method is only applicable to simulated data, where clean reference signals are available. Given a pair of enhanced and clean speech signals, we pass both through the SQA model and extract their corresponding internal features. The feature-based loss is defined as the L1 distance between the hidden representations of the enhanced and clean signals:
\begin{equation}
\mathcal{L}_{\text{feat}} = \left\| \mathbf{h}(\tilde{\mathbf{x}}_{\text{simu}}) - \mathbf{h}(\hat{\mathbf{x}}_{\text{simu}}) \right\|_1
\end{equation}
where \( \tilde{\mathbf{x}}_{\text{simu}} \) and \( \hat{\mathbf{x}}_{\text{simu}} \) denote the enhanced and clean reference signals, respectively, and \( \mathbf{h}(\cdot) \) denotes the hidden feature extracted from the SQA model. It is used for training on simulated data as shown in Figure~\ref{fig:overview}.

\begin{table*}[ht]
\renewcommand{\arraystretch}{1.1}  % increases line height
\centering
\caption{Comparison of metrics in the Uni-VERSA and Uni-VERSA-Ext, with optimization direction~(Opt.) and reference types.}
\vspace{-5pt}
\begin{tabular}{rl|l|c|c|c|c|l}
\hline
\textbf{\#} & \textbf{Domain} & \textbf{Metric} & \textbf{Range} & \textbf{Opt.} & \textbf{Uni-Versa~\cite{Uni-VERSA-Shi2025}} & \textbf{Uni-Versa-Ext} & \textbf{Reference Type} \\
\hline
1  & \multirow{10}{*}{Noise \& Distortion Level} & PESQ~\cite{PESQ-Rix2001}               & [1, 4.5]     & ↑ & \checkmark & \checkmark & Signal \\
2  & & DNSMOS~\cite{DNSMOS-Reddy2022}       & [1, 5]       & ↑ & \checkmark & \checkmark & No-reference \\
3  & & SIGMOS\_OVRL~\cite{SIGMOS-Ristea2025}       & [1, 5]       & ↑ &            & \checkmark & No-reference \\
4  & & SIGMOS\_NOISE~\cite{SIGMOS-Ristea2025}      & [1, 5]       & ↑ &            & \checkmark & No-reference \\
5  & & SIGMOS\_REVERB~\cite{SIGMOS-Ristea2025}     & [1, 5]       & ↑ &            & \checkmark & No-reference \\
6  & & SIGMOS\_COL~\cite{SIGMOS-Ristea2025}        & [1, 5]       & ↑ &            & \checkmark & No-reference \\
7  & & SIGMOS\_LOUD~\cite{SIGMOS-Ristea2025}       & [1, 5]       & ↑ &            & \checkmark & No-reference \\
8  & & SIGMOS\_SIG~\cite{SIGMOS-Ristea2025}        & [1, 5]       & ↑ &            & \checkmark & No-reference \\
9  & & LSD~\cite{LSD-Gray1976}                & [0, $\infty$) & ↓ &            & \checkmark & Signal \\
10 & & SI-SNR                & ($-\infty$, $\infty$) & ↑ & \checkmark &  & Signal \\
11 & & SDR~\cite{SDR-Vincent2006}                & ($-\infty$, $\infty$) & ↑ &  & \checkmark & Signal \\
\hline
12 & Prosody & F0-CORR & [--1, 1] & ↑ & \checkmark & & Signal \\
\hline
13 & \multirow{6}{*}{Naturalness}
  & MOS (human)        & [1, 5]       & ↑ & \checkmark & \checkmark & No-reference \\
14 & & UTMOS~\cite{UTMOS-Saeki2022}             & [1, 5]       & ↑ & \checkmark & \checkmark & No-reference \\
15 & & SHEET-base~\cite{SHEET-Huang2025}         & [1, 5]       & ↑ & \checkmark & & No-reference \\
16 & & Distill\_MOS~\cite{Distill_MOS-Stahl2025}       & [1, 5]       & ↑ &            & \checkmark & No-reference \\
17 & & NISQA\_MOS~\cite{NISQA-Mittag2021}         & [1, 5]       & ↑ &            & \checkmark & No-reference \\
18 & & SCOREQ~\cite{SCOREQ-Ragano2024}             & [1, 5]       & ↑ &            & \checkmark & No-reference \\
\hline
19 & \multirow{5}{*}{Intelligibility}
  & WER                & [0, $\infty$) & ↓ & \checkmark &  & Text \\
20 & & CER                & [0, $\infty$) & ↓ &            & \checkmark & Text \\
21 & & STOI              & [0, 1]        & ↑ &  \checkmark    & & Signal \\
22 & & ESTOI~\cite{ESTOI-Jensen2016}              & [0, 1]        & ↑ &            & \checkmark & Signal \\
23 & & SpeechBERTScore~\cite{SpeechBERTScore-Saeki2024}    & [0, 1]        & ↑ & \checkmark & \checkmark & Signal \\
24 & & PhonemeSimilarity~\cite{LPS-Pirklbauer2023}  & [0, 1] / [0, $\infty$) & ↑ & & \checkmark & Signal \\
\hline
25 & Speaker Characteristics & SpeakerSimilarity & [--1, 1] & ↑ & \checkmark & \checkmark & Signal \\
\hline
26 & Spectral Accuracy & MCD~\cite{MCD-Kubichek1993} & [0, $\infty$) & ↓ & & \checkmark & Signal \\
\hline
27 & Overall Quality & Ranking Score~\cite{URGENT-Zhang2024} & (0, 1) & ↓ & & \checkmark & Text \& Signal \\
\hline
\end{tabular}
\label{tab:universa-extended}
\vspace{-2em}
\end{table*}

\subsection{Fine-Tuning Supervised by SQA}

We now describe the full objective used for fine-tuning the speech enhancement model under SQA supervision. Depending on the data type—simulated or real-world—we apply different combinations of losses to balance fidelity, perceptual quality, and robustness.

\subsubsection{Simulated Data}

For samples with clean references, we use a combination of three losses. First, the \textit{multi-resolution spectrogram loss} \( \mathcal{L}_{\text{spec}} \) computes the L1 distance between the log-magnitude spectrograms of the enhanced and clean signals across multiple time-frequency resolutions:
\begin{equation}
\mathcal{L}_{\text{spec}} = \sum_{r} \left\| \log |\text{STFT}_r(\tilde{\mathbf{x}}_{\text{simu}})| - \log |\text{STFT}_r(\hat{\mathbf{x}}_{\text{simu}})| \right\|_1
\end{equation}
where $r$ denotes different STFT resolutions with varying window and hop sizes. It is used for training on real data as shown in Figure~\ref{fig:overview}.

Second, the \textit{score-based SQA loss} \( \mathcal{L}_{\text{score}} \) supervises the enhanced output using metric scores predicted by the Uni-VERSA-Ext model. Third, the \textit{feature-based SQA loss} \( \mathcal{L}_{\text{feat}} \) minimizes the discrepancy between SQA model features extracted from the enhanced and clean signals. The overall objective for simulated data is:
\begin{equation}
\mathcal{L}_{\text{simu}} = \lambda_{\text{spec}} \mathcal{L}_{\text{spec}} + \lambda_{\text{score}} \mathcal{L}_{\text{score}} + \lambda_{\text{feat}} \mathcal{L}_{\text{feat}}
\end{equation}

\subsubsection{Real-World Data}

For real recordings without clean references, we additionally introduce a \textit{regularization loss} \( \mathcal{L}_{\text{reg}} \) to prevent the SE model from drifting too far from its original domain and generating out-of-distribution outputs. \( \mathcal{L}_{\text{reg}} \) encourages the enhanced signal to remain close to the output of the initial SE model:
\begin{equation}
\mathcal{L}_{\text{reg}} = \sum_{r} \left\| \log |\text{STFT}_r(\tilde{\mathbf{x}}_{\text{real}})| - \log |\text{STFT}_r(\tilde{\mathbf{x}}_{\text{real}}^{(0)})| \right\|_1
\end{equation}
where \( \tilde{\mathbf{x}}_{\text{real}}^{(0)} \) and \( \tilde{\mathbf{x}}_{\text{real}} \) denote the enhanced real speech produced by the initial SE model and the SE model being fine-tuned, respectively. The total loss for real-world data is:
\begin{equation}
\mathcal{L}_{\text{real}} = \lambda_{\text{score}} \mathcal{L}_{\text{score}} + \lambda_{\text{reg}} \mathcal{L}_{\text{reg}}
\end{equation}

\section{Experiments}
\label{sec:exp}

\begin{table}[ht]
\centering
\caption{
Training data used for Uni-VERSA and Uni-VERSA-Ext. UG24/UG25 denote URGENT 2024/2025. "Source" refers to original recordings of the enhanced clips. \textbf{Note:} Although labeled as “test” sets, these are repurposed as training resources, as is standard in MOS prediction tasks.
}
\vspace{-5pt}
\resizebox{\linewidth}{!}{
\begin{tabular}{l|c|c|c}
\toprule
 & \# Clips & \# Source & Hours \\
\midrule
\multicolumn{4}{l}{\textbf{Uni-VERSA}} \\
\quad UG24 Non-blind Test & 9,690 & 850 & 249.05 \\
\midrule
\multicolumn{4}{l}{\textbf{Uni-VERSA-Ext}} \\
\quad UG24 Validation & 67,000 & 1,000 & 99.62 \\
\quad UG24 Non-blind Test\tablefootnote{The higher clip count in Uni-VERSA-Ext reflects the inclusion of valid submissions that were excluded from Uni-VERSA due to aborted evaluation.} & 153,900 & 850 & 285.16 \\
\quad UG24 Blind Test & 86,000 & 1,000 & 221.15 \\
\quad UG25 Validation & 8,000 & 1,000 & 19.42 \\
\quad UG25 Non-blind Test & 92,000 & 1,000 & 241.89 \\
\quad UG25 Blind Test & 67,500 & 900 & 121.40 \\
\bottomrule
\end{tabular}
}
\label{tab:universa-data}
\vspace{-2em}
\end{table}

\begin{table*}[h]
    \centering
    \caption{Comparison of Uni-VERSA and Uni-VERSA-Ext on the URGENT24 benchmark using commonly supported metrics. Results are shown as (LCC / SRCC).}
    \vspace{-6pt}
    \begin{tabular}{l|c|c|c|c|c|c|c}
    \toprule
        \multirow{2}{*}{Model} & \multicolumn{2}{c}{Noise \& Distortion Level} & \multicolumn{2}{|c|}{Naturalness} & Intelligibility & Speaker & \multirow{2}{*}{Average} \\
        \cmidrule{2-7}
         & PESQ & DNSMOS & MOS & UTMOS & SpeechBERTScore & SpeakerSimilarity & \\
        \midrule
        Uni-VERSA & 0.85 / 0.86 & 0.90 / \textbf{0.89} & 0.77 / 0.78 & 0.97 / 0.94 & 0.93 / 0.90 & 0.79 / 0.75 & 0.87 / 0.85 \\
        Uni-VERSA-Ext & \textbf{0.99} / \textbf{0.98}  & \textbf{0.95} / 0.82 & \textbf{0.95} / \textbf{0.90} & \textbf{0.99} / \textbf{0.97} & \textbf{0.97} / \textbf{0.92} & \textbf{0.96} / \textbf{0.95} & \textbf{0.97} / \textbf{0.92} \\
        \bottomrule
    \end{tabular}
    \label{tab:universa-comparison-common}
\vspace{-1em}
\end{table*}

\begin{table}[h]
    \renewcommand{\arraystretch}{1.3}  % increases line height
    \centering
    \vspace{-6pt}
    \caption{Out-of-domain correlation with MOS on CHiME-7 UDASE, BVCC, and BC2019. Results are reported as (LCC / SRCC).}
    \vspace{-5pt}
    \begin{tabular}{l|c|c|c}
    \toprule
     & CHiME-7 UDASE & BVCC & BC2019  \\
    \midrule
    Uni-VERSA & 0.68 / 0.66 & 0.68 / 0.66 & 0.65 / 0.56 \\
    Uni-VERSA-Ext & \textbf{0.69} / \textbf{0.77} & \textbf{0.82} / \textbf{0.81} & \textbf{0.79} / \textbf{0.78} \\
    \bottomrule
    \end{tabular}
    \label{tab:universa-ood}
\vspace{-2em}
\end{table}

\begin{table*}[ht]
    \centering
    \renewcommand{\arraystretch}{1.2}  % increases line height
    \caption{Comparison of supervision strategies using different loss configurations on DNS20 test sets, reported as (synthetic / synthetic reverb). LPS denotes Levenshtein Phoneme Similarity. Rank denotes the overall Ranking Score normalized to [0, 1]. Best results are in \textbf{bold}, based on full-precision values.}
    \vspace{-8pt}
    \resizebox {\linewidth} {!}{
    \begin{tabular}{r|l|c|cccccccc}
    \toprule
    \# & SQA Model & $\lambda_{\text{spec}}$/$\lambda_{\text{score}}$/$\lambda_{\text{feat}}$ & SDR~↑ & PESQ~↑ & LSD~↓ & MCD~↓ & LPS~↓ & ESTOI~↑ & SBERT~↑ & SpkSim~↑ \\
    \hline
    1 & \multicolumn{2}{c|}{Noisy} & 9.09 / 9.16 & 1.58 / 1.82 & 3.13 / 2.05 & 5.63 / 5.02 & 0.90 / 0.67 & 0.81 / \textbf{0.78} & 0.88 / 0.86 & \textbf{0.95} / \textbf{0.88} \\
    \hline
    2 & None & 1 / 0 / 0 & 18.21 / 10.97 & 2.86 / 1.99 & 2.49 / 2.37 & 2.03 / 3.99 & 0.96 / 0.83 & 0.93 / 0.77 & 0.93 / 0.88 & 0.93 / 0.84 \\
    \hline
    3 & Uni-VERSA & 1 / 1 / 1 & 18.29 / 10.75 & 2.91 / 1.88 & 2.45 / 2.33 & \textbf{1.98} / 3.99 & 0.96 / 0.83 & 0.93 / 0.76 & 0.93 / 0.87 & 0.93 / 0.83 \\
    \hline
    4 & \multirow{6}{*}{\shortstack[l]{Uni-VERSA-Ext\\(proposed)}} & 1 / 1 / 1 & 18.27 / 10.72 & 2.91 / 1.90 & 2.41 / 2.32 & 1.99 / 3.94 & 0.96 / 0.83 & 0.93 / 0.77 & 0.93 / 0.88 & 0.93 / 0.83 \\
    5 &  & 1 / 1 / 0 & \textbf{18.30} / 10.74 & \textbf{2.91} / 1.88 & 2.48 / 2.35 & 1.99 / 3.99 & 0.96 / \textbf{0.83} & 0.93 / 0.76 & 0.93 / 0.87 & 0.93 / 0.83 \\
    6 &  & 1 / 0 / 1 & 18.25 / 10.80 & 2.91 / 1.91 & \textbf{2.41} / \textbf{2.31} & 1.99 / \textbf{3.91} & 0.96 / 0.83 & 0.93 / 0.77 & 0.93 / 0.88 & 0.93 / 0.84 \\
    7 &  & 0 / 1 / 1 & 17.93 / \textbf{11.15} & 2.90 / \textbf{1.95} & 2.50 / 2.46 & 2.20 / 3.98 & \textbf{0.96} / 0.83 & \textbf{0.93} / 0.77 & \textbf{0.93} / \textbf{0.88} & 0.92 / 0.82 \\
    8 &  & 0 / 1 / 0 & 1.56 / -1.43 & 1.04 / 1.03 & 5.71 / 5.84 & 7.64 / 9.51 & 0.29 / 0.15 & 0.33 / 0.18 & 0.59 / 0.52 & 0.19 / 0.08 \\
    9 &  & 0 / 0 / 1 & 17.94 / 11.12 & 2.90 / 1.95 & 2.49 / 2.46 & 2.19 / 3.99 & 0.96 / 0.83 & 0.93 / 0.77 & 0.93 / 0.88 & 0.92 / 0.83 \\
    \hline %\addlinespace[3pt]
    \multicolumn{3}{r|}{(continued)} & CER~↓ & UTMOS~↑ & SIGMOS.OVRL~↑ & SCOREQ~↑ & DNSMOS~↑ & Distill-MOS~↑ & NISQA~↑ & \cellcolor[HTML]{EEEEEE}\textbf{Rank~\cite{URGENT-Zhang2024}~↓} \\
    \hline
    1 & \multicolumn{2}{c|}{Noisy} & \textbf{0.25} / 0.26 & 2.37 / 1.30 & 2.32 / 2.07 & 2.79 / 2.10 & 2.48 / 1.39 & 3.06 / 2.32 & 2.58 / 1.66 & \cellcolor[HTML]{EEEEEE}0.60 / 0.48\\
    \hline
    2 & None & 1 / 0 / 0 & 0.34 / 0.27 & 3.60 / 1.38 & 3.06 / 2.42 & 3.98 / 2.53 & 3.25 / 2.59 & 4.07 / 2.74 & 4.25 / 2.36 & \cellcolor[HTML]{EEEEEE}0.42 / 0.40 \\
    \hline
    3 & Uni-VERSA & 1 / 1 / 1 & 0.32 / 0.26 & 3.61 / 1.38 & 3.11 / 2.46 & 4.03 / 2.51 & 3.24 / 2.52 & 4.09 / 2.65 & 4.26 / 2.34 & \cellcolor[HTML]{EEEEEE}0.35 / 0.48 \\
    \hline
    4 & \multirow{6}{*}{\shortstack[l]{Uni-VERSA-Ext\\(proposed)}} & 1 / 1 / 1 & 0.31 / 0.25 & 3.69 / 1.39 & 3.00 / 2.50 & 4.09 / 2.69 & 3.28 / 2.61 & 4.29 / 2.67 & 4.49 / 2.72 & \cellcolor[HTML]{EEEEEE}\textcolor{red}{\textbf{0.33}} / 0.42 \\
    5 &  & 1 / 1 / 0 & 0.33 / 0.25 & 3.61 / 1.38 & 3.12 / 2.47 & 4.03 / 2.50 & 3.25 / 2.52 & 4.08 / 2.62 & 4.29 / 2.36 & \cellcolor[HTML]{EEEEEE}0.38 / 0.51 \\
    6 &  & 1 / 0 / 1 & 0.32 / 0.22 & 3.69 / 1.39 & 3.01 / 2.52 & 4.09 / 2.69 & 3.26 / 2.58 & 4.22 / 2.64 & 4.47 / 2.71 & \cellcolor[HTML]{EEEEEE}0.34 / 0.38 \\
    7 &  & 0 / 1 / 1 & 0.31 / 0.24 & 3.84 / 1.42 & \textbf{3.14} / 2.58 & \textbf{4.21} / \textbf{2.85} & \textbf{3.33} / \textbf{2.77} & \textbf{4.46} / \textbf{2.76} & \textbf{4.59} / \textbf{2.94} & \cellcolor[HTML]{EEEEEE}0.39 / \textcolor{red}{\textbf{0.38}} \\
    8 &  & 0 / 1 / 0 & 0.77 / 1.05 & 2.34 / \textbf{1.64} & 2.47 / 2.17 & 3.07 / 2.66 & 2.75 / 2.35 & 2.77 / 2.24 & 2.75 / 1.93 & \cellcolor[HTML]{EEEEEE}0.77 / 0.76 \\
    9 &  & 0 / 0 / 1 & 0.32 / \textbf{0.23} & \textbf{3.84} / 1.42 & 3.13 / \textbf{2.59} & 4.21 / 2.84 & 3.33 / 2.76 & 4.41 / 2.69 & 4.56 / 2.92 & \cellcolor[HTML]{EEEEEE}0.39 / 0.38 \\
    \bottomrule
    \end{tabular}
    }
    \label{tab:supervision-ablation}
\vspace{-.5em}
\end{table*}

\subsection{Datasets}

\paragraph{Uni-VERSA-Ext} The dataset used for Uni-VERSA-Ext significantly extends that of the original Uni-VERSA model. While Uni-VERSA was trained on system submissions from the URGENT 2024 blind test set, Uni-VERSA-Ext incorporates all available submissions from both the URGENT 2024 and URGENT 2025 challenges. This expanded dataset introduces greater diversity across source utterances, enhancement methods, and acoustic conditions, and includes more multilingual and MOS-labeled samples from URGENT 2025. A detailed breakdown of the training data is provided in Table~\ref{tab:universa-data}. 
\textbf{Note:} Although the URGENT challenge datasets refer to these sets as “test” sets (e.g., blind/non-blind test), they are repurposed as training resources following common practice in MOS prediction challenges.
For in-domain evaluation, we follow~\cite{Uni-VERSA-Shi2025} and use a subset of the URGENT 2024 non-blind test set, which comprises 4,200 enhanced utterances derived from 50 source recordings. For out-of-domain evaluation, we include the CHiME-7 UDASE~\cite{CHiME-7-UDASE-Leglaive2023} set for SE, as well as BVCC~\cite{BVCC-Cooper2022} and BC19~\cite{BC19-Wu2019} for evaluating voice conversion~(VC) and text-to-speech~(TTS) systems. The out-of-domain test sets are prepared using the SHEET recipe~\cite{SHEET-Huang2024}\footnote{\url{https://github.com/unilight/sheet}}.  We adopt linear correlation coefficient (LCC) and Spearman rank correlation coefficient (SRCC) as evaluation metrics.

\paragraph{Speech Enhancement} We train the SE model using both simulated and real-world data. For simulated data, we follow the URGENT 2025 Challenge setup, where 2,500 hours of clean speech are dynamically mixed with 400 hours of noise recordings. Unless otherwise specified, all experiments use a randomly sampled 100-hour subset to ensure faster convergence. For completeness, we also report results using 700-hour and full 2,500-hour training sets. For real data, we use single-channel recordings from the CHiME-4 dataset. Evaluation is conducted on two representative benchmarks: the non-blind test set from the DNS-2020 Challenge~\cite{DNS2020-Reddy2020} and the CHiME-4 dataset~\cite{Analysis-Vincent2017}. Note that the \textsc{synthetic} and \textsc{synthetic reverb} test sets in DNS-2020 do not provide transcripts; we manually annotated them for ASR evaluation~\footnote{\url{https://github.com/Emrys365/DNS_text}}.

\subsection{Models and Training Details}

\paragraph{Uni-VERSA-Ext} We adopt the reference-free and text-free variant of the Uni-VERSA architecture in~\cite{Uni-VERSA-Shi2025}. Our implementation follows the same setup using ESPnet~\cite{ESPnet-watanabe2018}. Acoustic features are extracted from a frozen WavLM-Large model~\cite{WavLM-Chen2022} via a learnable layer-wise weighted summation implemented with S3PRL~\cite{S3PRL-Yang2021}. The model comprises three parallel encoders, each consisting of 4 Transformer layers with 4-head self-attention (256-dimensional), a 1024-dimensional feedforward network, and a dropout rate of 0.1. A mean pooling operation is applied to the output of each encoder, followed by a linear projection to predict metric scores. 
We train the model for 20 epochs using a batch size of 16. 
The optimizer is AdamW with an initial learning rate of 0.001 and a linear warm-up schedule over 25{,}000 steps.

\paragraph{Speech Enhancement} We use BSRNN~\cite{BSRNN-Luo2022} as the backbone of our SE model, initialized with weights pretrained on the URGENT 2025 Challenge training set. All audio is resampled to 48kHz before processing. The model operates in the time-frequency domain using an STFT-based encoder and decoder, each configured with a 960-point FFT and a hop size of 480. The encoded spectrogram is processed by a BSRNN separator composed of 6 bidirectional LSTM layers with 196 hidden channels. Fine-tuning is performed using the Adam optimizer with an initial learning rate of $1\times10^{-5}$, a StepLR learning rate scheduler, and a batch size of 8. We fine-tune the model for 50{,}000 steps on the target datasets.

\subsection{Performance Evaluation of Uni-VERSA-Ext}

As shown in Table~\ref{tab:universa-comparison-common}, Uni-VERSA-Ext achieves consistent improvements across all metric categories in the in-domain setting. In particular, the correlation with human-rated MOS improves substantially, suggesting that the extended model better captures perceptual quality as judged by human listeners. In the out-of-domain setting (Table~\ref{tab:universa-ood}), Uni-VERSA-Ext outperforms Uni-VERSA on all three benchmark datasets, achieving higher LCC and SRCC scores. These results demonstrate that the proposed extensions lead to better generalization across different tasks and conditions.

\subsection{Comparison of supervision approaches}

\begin{figure}[htbp]
\vspace{-10pt}
\centerline{\includegraphics[width=\linewidth]{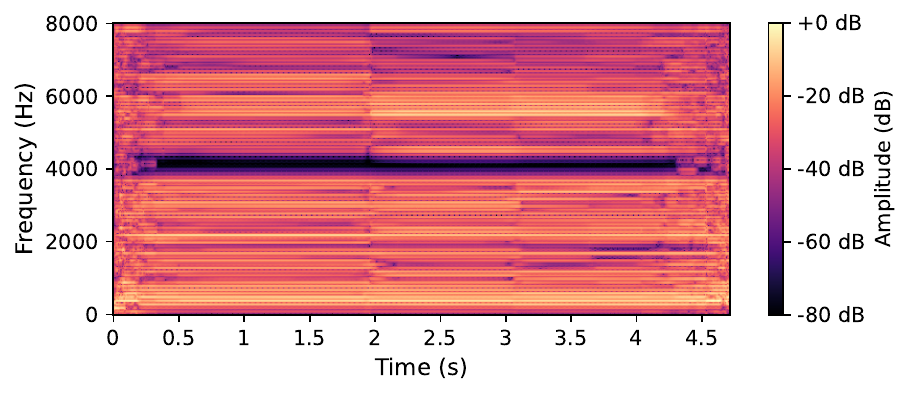}}
\vspace{-10pt}
\caption{An example of an adversarial sample resulting from training with $\mathcal{L}_{\text{score}}$ only. Although the audio is clearly nonspeech, Uni-VERSA predicts it as high quality (e.g., PESQ: 4.77; UTMOS: 4.02; SBERTScore: 0.98).}
\label{fig:adv_spec}
% \vspace{-1em}
\end{figure}

\begin{table}[h]
    \centering
    \caption{
Comparison of supervision strategies on the CHiME-4 et05 real test set. Only non-intrusive metrics are reported. "Baseline" refers to evaluation using the initial SE model.
}
    \vspace{-5pt}
    \resizebox {\linewidth} {!}{
    \begin{tabular}{c|cccccc}
    \toprule
    $\lambda_{\text{score}}$ / $\lambda_{\text{reg}}$ & 
    DNS. & UT. & NISQA & SIG.OVRL & Distill. & CER \\
    \midrule
    noisy &  1.46 & 1.41 & 1.20 & 1.93 & 2.12 & 7.98 \\
    \midrule
    baseline &  2.71 & 2.65 & 3.02 & 2.38 & 2.60 & 8.13 \\
    \midrule
    1 / 1 &  2.70 & 2.66 & 3.06 & 2.38 & 2.59 & 7.95 \\
    1 / 0 &  \multicolumn{6}{c}{(diverge)} \\
    \bottomrule
    \end{tabular}
    }
    \label{tab:real-supervision-ablation}
    \vspace{-1em}
\end{table}

\begin{table}[h]
    \centering
    \caption{Effect of simulated data scale on CHiME-4 et05 real test set. Only non-intrusive metrics are reported for real recordings.}
    \vspace{-5pt}
    \resizebox {\linewidth} {!}{
    \begin{tabular}{l|c|ccccccc}
    \toprule
    Data  & \multirow{2}{*}{SQA} & \multirow{2}{*}{DNS.} & \multirow{2}{*}{UT.} & \multirow{2}{*}{NISQA} & \multirow{2}{*}{SIG.OVRL} & \multirow{2}{*}{Distill.} & \multirow{2}{*}{CER} \\
    (hrs.)  &  & & & & &  &  \\
    \midrule
    \multicolumn{2}{c|}{noisy} &  1.46 & 1.41 & 1.20 & 1.93 & 2.12 & 7.98 \\
    \midrule
    \multirow{2}{*}{100} & \xmark &  2.71 & 2.65 & 3.02 & 2.38 & 2.60 & 8.13  \\
     & \cmark &  \textbf{2.97} & \textbf{3.35} & \textbf{4.18} & \textbf{2.85} & \textbf{3.60} & \textbf{7.79} \\
     \midrule
     \multirow{2}{*}{700} & \xmark & 2.75 & 2.88 & 3.30 & 2.58 & 2.93 & \textbf{6.88} \\
     & \cmark  & \textbf{2.83} & \textbf{3.04} & \textbf{4.05} & \textbf{3.32} & \textbf{3.24} & 7.81 \\
     \midrule
     \multirow{2}{*}{2500} & \xmark & 2.95 & 3.28 & 3.92 & \textbf{2.87} & 3.34 &  \textbf{6.67} \\
     & \cmark  & \textbf{2.97} & \textbf{3.35} & \textbf{4.18} & 2.85 & \textbf{3.60} & 7.79 \\
    \bottomrule
    \end{tabular}
    }
    \label{tab:training-data-comparison}
\vspace{-2em}
\end{table}

Table~\ref{tab:supervision-ablation} investigates the impact of different loss combinations on model performance. All models are fine-tuned on a randomly selected 100-hour subset of the URGENT 2025 challenge training data, using checkpoints pre-trained on the same subset. Notably, line 4 and line 7, which apply both feature and score level SQA guidance, achieve the best overall performance as reflected in the ranking score column.

Training with only $\mathcal{L}_{\text{score}}$ (line 7) results in the worst performance. This is likely due to the vulnerability of neural networks to adversarial attacks~\cite{Adv-madry2018,Adv-bhanushali2024}. When guided solely by gradients from the SQA model, the SE network can easily figure out a simple shortcut that leads to non-speech outputs with high SQA scores. As illustrated in Figure~\ref{fig:adv_spec}, the model can learn to generate non-speech signals that are misjudged as high-quality by the SQA model.

Interestingly, comparing lines 6 and 9, we find that incorporating $\mathcal{L}_{\text{spec}}$ leads to improved performance on intrusive metrics (which rely on clean reference signals, such as SDR), but results in slightly worse outcomes on non-intrusive metrics (e.g., Distill MOS). This observation aligns with the intuition that optimizing for intrusive criteria primarily enhances metrics that depend on reference-based comparisons.

Finally, CER performance degrades on the \textsc{synthetic} set but improves on the \textsc{synthetic reverb} set, as denoising may introduce artifacts harmful to ASR. But dereverberation addresses the more detrimental effect of reverberation. As shown in Table~\ref{tab:training-data-comparison}, the degradation on the \textsc{synthetic} set can be mitigated with larger-scale training data.

\subsection{Effect of SQA supervision with real data}

As shown in Table~\ref{tab:real-supervision-ablation}, fine-tuning on real-world data alone fails to converge, likely due to the adversarial attack phenomenon discussed earlier. Incorporating the proposed $\mathcal{L}_{\text{reg}}$ mitigates this issue and leads to improved performance. Table~\ref{tab:training-data-comparison} presents results from fine-tuning pretrained SE models on mixed datasets, where varying amounts of simulated data are combined with the CHiME-4 real training set. The experiments show consistent improvements across all perceptual metrics. However, ASR performance, measured by CER, degrades in the 700-hour and 2500-hour settings. This may be due to a conflict between optimizing for perceptual quality and ASR objectives. Additionally, CER, unlike the reference-free perceptual metric, is inherently more challenging to optimize, as it depends on accurate text recognition.

\section{Conclusions}
\label{sec:conclusions}
In this work, we propose a novel framework that leverages a pretrained SQA model to guide the fine-tuning of a SE model. We first extend the Uni-VERSA model to Uni-VERSA-Ext by incorporating additional training data and expanding the set of predicted quality metrics. We then use Uni-VERSA-Ext to provide gradient-based supervision for fine-tuning a pretrained SE model. Experiments on SQA benchmarks demonstrate that Uni-VERSA-Ext achieves stronger correlation with perceptual quality metrics compared to Uni-VERSA. Furthermore, evaluations on both real and simulated SE datasets show that SQA-guided SE training leads to improved performance across a range of perceptual metrics.

\section*{Acknowledgment}
This work was supported in part by China STI 2030-Major Projects under Grant No. 2021ZD0201500, in part by Shanghai Municipal Science and Technology Commission Project under Grant 2021SHZDZX0102, and in part by the the Emerging Interdisciplinary Research Specialized Project of Shanghai Municipal Health Commission (No. 2022JC024).

% This work was supported by ...
\newpage
\bibliographystyle{IEEEtran}
\bibliography{ref}

\end{document}